\documentclass[twocolumn,showpacs,amsmath,amssymb,amsfonts,nofootinbib,prl]
{revtex4-1}
\usepackage{graphicx}
\usepackage{dcolumn}
\usepackage{bm}
\usepackage{ulem}
\usepackage{overpic}

\def\beq{\begin{equation}}
\def\eeq{\end{equation}}

\begin{document}
\input{epsf}

\title{Soft gluon emission off a heavy quark revisited}
\author{Raktim Abir$^1$, Carsten Greiner$^2$, Mauricio Martinez$^3$, 
Munshi G. Mustafa$^{1}$, and Jan Uphoff$^2$}
\affiliation{$^1$Theory Division, Saha Institute of Nuclear Physics  
1/AF Bidhannagar, Kolkata 700064, India.}

\affiliation{$^2$Institute f\"ur Theoretische Physik, Johann Wolfgang Goethe
University, Max-von-Laue-Strasee 1, D-60438 Frankfurt, Germany} 

\affiliation{$^3$Departamento de F\'isica de Part\'iculas, Universidade de Santiago de Compostela,
E-15782 Santiago de Compostela, Galicia, Spain}

\begin{abstract}
An improved generalized suppression factor for gluon emission off a heavy quark is derived within perturbative QCD, which is valid for the full range of rapidity of the radiated gluon and also has no restriction on the scaled mass of the quark with its energy. In the appropriate limit it correctly reproduces the usual dead cone factor in the forward rapidity region.  On the other hand, this improved suppression factor becomes close to unity in the backward direction. This indicates a small suppression of gluon emission in the backward region, which should have an impact on the phenomenology of heavy quark energy loss in the hot and dense matter produced in ultra relativistic heavy ion collisions.
\end{abstract}


\date{\today}
\maketitle
The main aim of ongoing ultra relativistic heavy ion collisions is to study the properties of nuclear or hadronic matter at extreme conditions. A particular goal lies in the identification of a new state of matter formed in such collisions, the quark-gluon plasma (QGP), where the quarks and gluons are deconfined from the nucleons and move freely over an extended space-time region. Various measurements taken at CERN Super Proton Synchrotron (SPS)~\cite{heinz} and BNL Relativistic
Heavy Ion Collider (RHIC)~\cite{white,dilep,phot,ellip,jet,phenix,non-photonic} lead to a wealth of information for the formation of the QGP during the first several fm/c of the collisions through the hadronic final states.  New data from experiments at CERN Large Hadron Collider (LHC)~\cite{lhc} have supported the existence of such a state of matter.

Some of the important features of the plasma produced in heavy ion collisions include energy loss and jet quenching of high energetic partons, viz., light and heavy quarks. The Gunion-Bertsch (GB) formula \cite{gunion82} for gluon emission from the processes $qq\rightarrow qqg$ has been widely used in different phenomenological studies of heavy ion collisions, in particular for radiative energy loss of high energy partons propagating through a thermalized QGP~\cite{Fochler:2010wn,mgm02,dokshit01,dead,dead1,alam10,kope10,Fochler:2008ts,wong96,Abir10,das10,mgm08,Gossiaux10,wicks07}. The energy loss is presently a field of high interest in view of jet quenching of high energy partons, viz., both light~\cite{mgm08,horowitz10,majum10,renk10} and heavy quarks~\cite{Fochler:2010wn,mgm02,dokshit01,dead,alam10,kope10,Fochler:2008ts,jamil10,armesto,Uphoff11a,Uphoff11b,Gossiaux10,wicks07,horowitz10,Zhang04}. Generally, one expects that jet quenching for heavy quarks should be weaker than that of light quarks. In contrast the non-photonic data at RHIC~\cite{non-photonic} reveal a similar suppression for heavy flavored hadrons compared to that of light hadrons.

An early attempt to calculate the heavy quark energy loss in a QGP medium was done in Ref.~\cite{mgm02} by using the GB formula of gluon emission for light quark scattering~\cite{gunion82} and just modifying the relevant kinematics for heavy quarks. 
Later the soft gluon emission formula for heavy quarks in the high energy approximation~\cite{dead} was renewed in
Ref.~\cite{dokshit01} for the small angle limit. Soft gluon emission
from a heavy quark was found to be suppressed in the forward direction 
compared to that from a light quark due to the mass effect (dead cone
effect). The corresponding suppression factor was obtained as~\cite{dokshit01},
\begin{eqnarray}
\left(1+\frac{\theta_0^2}{\theta^2}\right)^{-2}\,,
\label{old_dead_cone}
\end{eqnarray}
where $\theta_0=M/E \ll 1$. $E$ is the energy of the heavy quark with mass, $M$ and $\theta$, the
angle between the heavy quark and the radiated gluon. 
Often in the literature \cite{dead1,alam10} the expression
\begin{equation}
\frac{k_\bot^4 }{ ( k_\bot^2 + \theta_0^2 \omega^2 )^2 }
= \left (1+\frac{\theta_0^2}{\sin^2\theta}\right )^{-2}
\label{old_dead_cone_sin2}
\end{equation}
is used as the suppression factor to calculate heavy quark energy loss in heavy ion collisions.
Here, $k_\bot$ denotes the transverse momentum of the emitted gluon and is related to its energy $\omega$ by $k_\bot=\omega \sin\theta $.
For small angles \eqref{old_dead_cone_sin2} reduces to \eqref{old_dead_cone}.
However, \eqref{old_dead_cone_sin2} produces not only a 
dead cone in forward direction $(\theta  \ll 1)$ but also in the backward region $(\theta \sim {\pm \pi})$. The uniform use of such a dead cone for heavy quark energy loss may not be accurate 
enough since high energy scatterings off partons are associated with gluon emission in 
all directions with varying magnitude~\cite{dead}, as we will see below. 

In this article we revisit the issue and make an attempt to generalize the gluon emission off a heavy quark by relaxing the constrains imposed in earlier calculations on the emission angle of the radiated gluons and the scaled mass of the heavy quark with its energy. 
We have found a generalized expression of the suppression factor that is identical to \eqref{old_dead_cone} and \eqref{old_dead_cone_sin2} for large $E$ and small $\theta$ but unlike \eqref{old_dead_cone_sin2} smoothly becomes unity (no suppression) in the backward direction. This supports the point of
Ref.~\cite{dokshit01} that the main modification of the gluon radiation spectrum due to a non-zero quark mass occurs at 
small angles (forward direction) and not at large angles (backward direction). 

\begin{figure}[t]
\centering
\begin{overpic}[width=0.8\linewidth]{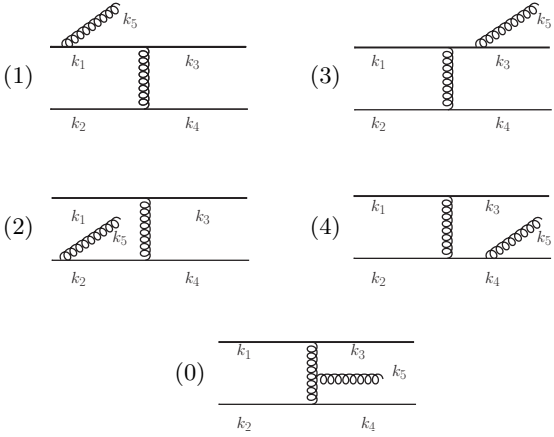}
\put(-8,67){(1)}
\put(-8,38){(2)}
\put(51,67){(3)}
\put(51,38){(4)}
\put(25,10){(0)}
\end{overpic}
\caption{Five tree level Feynman diagrams for the process $Qq\rightarrow Qqg$. 
In each diagram the thick upper line represents the heavy quark ($Q$) whereas the 
thin lower line represents the background light quark.}
\label{diagram}
\end{figure}
In Fig.~\ref{diagram} the five Feynman diagrams for the process $Qq\rightarrow Qqg$ are shown. 
According to the notation used in the figure, the Mandelstam variables are
\begin{subequations}
\label{Mandelstam} 
\begin{align}
&\hspace{-0.3cm}s=(k_1+k_2)^2\,,\hspace{0.3cm} 
s^\prime=(k_3+k_4)^2\,,\\
&\hspace{-0.3cm}u=(k_1-k_4)^2\,,\hspace{0.3cm} 
u^\prime=(k_2-k_3)^2\,, \\
&\hspace{-0.3cm}t=(k_1-k_3)^2\,,\hspace{0.3cm} 
t^\prime=(k_2-k_4)^2\,, 
\end{align}
\end{subequations}
with
\begin{eqnarray}
 &&\hspace{-0.4cm} s+t+u+s'+t'+u'=4M^2\ .
\end{eqnarray}
Soft gluon emission ($k_5\rightarrow 0$)~\cite{berends81} implies
$t^\prime \rightarrow t, \ \ s^\prime \rightarrow s, \ \ u^\prime \rightarrow u$. In the center of  momentum frame we consider the case where the energy of the emitted gluon, $\omega$ is much smaller than the momentum transfer $\sqrt{|t|} \approx q_\bot$ from the projectile (heavy quark) to the target (light quark) which again is small compared to the energy of heavy quark $E$. This leads to the hierarchy
\begin{eqnarray}
E \gg \sqrt{|t|} \gg \omega \,.
\label{hierarchy}
\end{eqnarray}
It is important to note that the scaled mass of the heavy quark with its energy $M/E$ and the gluon emission angle $\theta$ are free from any constrain. 

The gauge invariant amplitude for the process $Qq\rightarrow Qqg$ can be written as the squared matrix elements from the diagrams of Fig.~\ref{diagram}, including their interference terms, 
\begin{equation}
\left | {\cal M}_{Qq\rightarrow Qqg}\right |^2 = \sum_{i\ge j} 
{\cal M}_{ij}^{ 2} \ , \label{matrix_element_def}
\end{equation}
where $i$ and $j$ run from $0$ to $4$ and ${\cal M}_{ij}^{ 2} = {\cal M}_{i} {\cal M}_{j}^*$ with ${\cal M}_{i}$ being the matrix element of diagram $i$ (see Fig.~\ref{diagram}).

With the hierarchy indicated in \eqref{hierarchy} the different matrix elements squared are obtained in the Feynman gauge as
\begin{eqnarray}
\label{subamplitude}
{\cal M_{\rm 11}^{\rm 2}}&=&{\cal M_{\rm 33}^{\rm 2}} =\frac{128}{27}
g^6\frac{s^2}{t^2}\frac{1}{k_\bot^2}\left[\frac{M^2}{s}-1+{\cal J}\right]{\cal J} , \nonumber \\
{\cal M_{\rm 00}^{\rm 2}}&=&{\cal M_{\rm 22}^{\rm 2}}={\cal M_{\rm 44}^{\rm 2}}= 0, \nonumber \\
{\cal M_{\rm 13}^{\rm 2}} &=& \frac{128}{27}g^6
\frac{s^2}{t^2}  
\frac{1}{k_\bot^2}\left[\frac{1}{4} \left( \frac{M^2}{s}-1+{\cal J}\right)\right]{\cal J} , \nonumber \\
{\cal M_{\rm 14}^{\rm 2}}&=&{\cal M_{\rm 23}^{\rm 2}} = 
\frac{128}{27}g^6\frac{s^2}{t^2}\frac{1}{k_\bot^2}\left[\frac{7}{8}\left(1-\frac{M^2}{s}\right)\right]{\cal J}, \nonumber\\
{\cal M_{\rm 12}^{\rm 2}}&=&{\cal M_{\rm 34}^{\rm 2}} = \frac{128}{27}g^6\frac{s^2}{t^2}\frac{1}{k_\bot^2}\left[\frac{1}{4}\left(1-\frac{M^2}{s}\right)\right]{\cal J}, \nonumber \\
{\cal M_{\rm 24}^{\rm 2}} &=& {\cal M_{\rm 10}^{\rm 2}} = {\cal M_{\rm 20}^{\rm 2}} = {\cal M_{\rm 30}^{\rm 2}} = {\cal M_{\rm 40}^{\rm 2}} = 0 , 
\end{eqnarray}
with
\begin{eqnarray}
{\cal J}&=&1-\left[\left(\frac{s}{M^2}-1\right)\sin^2(\theta/2)+1\right ]^{-1}~.
\end{eqnarray}
For the calculation of the matrix elements we used REDUCE and CalcHep programs~\cite{Semenov09}. We note that the results are in accordance with the matrix elements obtained from the process 
$q{\bar q}\rightarrow Q{\bar Q}g$~\cite{Kunszt80} by crossing the light antiquark and the heavy antiquark. Also, by proper transformation to light-cone variables, this result reduces to that obtained in Ref.~\cite{Gossiaux10} within scalar QCD approximation and in light-cone gauge. 

The gauge invariant amplitude for the process $Qq\rightarrow Qqg$ can now be obtained by summing all the sub-amplitudes~(\ref{subamplitude}),
\begin{eqnarray}
\left|{\cal M}_{Qq\rightarrow Qqg}\right|^2  
&=&12g^2\left|{\cal M}_{Qq\rightarrow Qq}\right|^2\frac{1}{k_\bot^2}\frac{{\cal J}^{2}}{\left(1-\frac{M^2}{s}\right)^{2}}    \nonumber \\
&=& 12g^2\left|{\cal M}_{Qq\rightarrow Qq}\right|^2\frac{1}{k_\bot^2}\left(1+\frac{M^2}{s\tan^{2}(\frac{\theta}{2})}\right)^{-2} \nonumber \\
&=& 12g^2\left|{\cal M}_{Qq\rightarrow Qq}\right|^2\frac{1}{k_\bot^2}\left(1+\frac{M^2}{s}e^{2\eta}\right)^{-2}\,,
\label{GB_formula_with_new_dead_cone}
\end{eqnarray}
where $\eta =-\ln[ \tan (\theta/2)]$, the rapidity of the emitted massless gluon. The two body amplitude is given by
\begin{equation}
\left|{\cal M}_{Qq\rightarrow Qq}\right|^2=\frac{8}{9}g^4\frac{s^2}{t^2} \left(1-\frac{M^2}{s}\right)^2\,.
\label{matrix_element_22}
\end{equation}
Equation~(\ref{GB_formula_with_new_dead_cone}), which is the main result of the present article, carries a generalized suppression factor, ${\cal D}$ as 
\begin{eqnarray}
{\cal D}&=& \left(1+\frac{M^2}{s\tan^{2}(\frac{\theta}{2})}\right)^{-2}\,.
\label{new_dead_cone}
\end{eqnarray}
This improved suppression factor is valid in the full range of $\theta$ (or
rapidity of the emitted gluon)  (i.e., $-\pi<\theta<+\pi$) and in the full
range of $M/{\sqrt s}$ (i.e., $0<M/{\sqrt s}<1$) as compared to 
Ref.~\cite{dead}.
As a note, the relation between the center of mass energy $\sqrt s$ and the energy of the heavy quark $E$ reads
\begin{equation}
s = 2E^2 + 2E \sqrt{E^2-M^2} - M^2 \ .
\end{equation}

Below we discuss our results in more detail. First we consider two limits:
\begin{enumerate}
\item {\it Gunion-Bertsch limit}: For $M = 0$, (\ref{GB_formula_with_new_dead_cone}) reduces to the well known result of Gunion and Bertsch~\cite{gunion82} as
\begin{eqnarray}
\left| {\cal M}_{qq' \rightarrow qq'g} \right|^2&=&12g^{2}\left| 
{\cal M}_{qq' \rightarrow qq'} \right|^2 \frac{1}{k_\perp^{2}}  \nonumber \\
&\simeq&12g^{2}\left| 
{\cal M}_{qq' \rightarrow qq'} \right|^2 \frac{1}{k_\perp^{2}}\frac{q_\bot^2}{(q_\bot-k_\bot)^2} \nonumber \\
&=& \left| {\cal M}_{qq' \rightarrow qq'g} \right|^2_{\rm GB} ,
\end{eqnarray}
where we have used \eqref{hierarchy} that implies $q_\perp \gg k_\perp$ \cite{wong96,Abir10,das10}.

\item {\it Dokshitzer and Kharzeev's result}: 
In the limit $M \ll \sqrt{s}$ and $\theta  \ll 1$, it is $\sqrt{s} \simeq 2E$ and $\tan(\theta/2)\simeq \theta/2$ and (\ref{GB_formula_with_new_dead_cone}) reduces to
\begin{eqnarray}
\left| {\cal M}_{Qq\rightarrow Qqg} \right|^2&=&12g^{2}\left| 
{\cal M}_{Qq\rightarrow Qq} \right|^2 \frac{1}{k_\perp^{2}} \left(1+\frac{M^2}{E^2\theta^2}\right)^{-2} \nonumber \\
&\simeq &12g^{2}\left| 
{\cal M}_{Qq\rightarrow Qq} \right|^2 \frac{1}{k_\perp^{2}}\left(1+\frac{\theta_0^2}{\theta^2}\right)^{-2} \,,
\end{eqnarray}
where $\theta_0=M/E$. This expression is precisely the result derived in Ref.~\cite{dokshit01}.
\end{enumerate}

For convenience, we define ${\cal R}$ as the ratio of the squared matrix element of the $2\rightarrow3$ to that of the $2\rightarrow2$ processes,
\begin{eqnarray}
{\cal R}=\frac{\left| {\cal M}_{Qq \rightarrow Qqg}\right|^2}{\left|{\cal M}_{Qq\rightarrow Qq}\right|^2}= 3g^2\frac{1}{\omega^2}\left(\frac{e^{\eta}+e^{-\eta}}{1+\frac{M^2}{s}e^{2\eta}}\right)^{2} \label{ratio_R} \,.
\end{eqnarray}
We note that this ratio, ${\cal R}$ is related to the gluon emission multiplicity distribution~\cite{wong96,Abir10,das10} as
$dn_g/d\eta dk_\bot^2={\cal R}/16\pi^3$.
For the massless case, ${\cal R}^{M\rightarrow 0}$ is symmetric in rapidity. In contrast, a finite mass of the quark renders ${\cal R}$ to be asymmetric in rapidity. To explore this in more detail we consider the following rapidity regions:
\begin{enumerate}

\item {\it Forward rapidity $(\eta \gg 0)$}: In this case \eqref{ratio_R} reduces to
\begin{eqnarray}
{\cal R}_{\eta \gg 0}
&\rightarrow & 3g^2\frac{1}{\omega^2}\frac{s^2}{M^4}e^{-2\eta} \,.
\end{eqnarray}
Clearly, in this region of rapidity the gluon emission is exponentially suppressed, which indicates the presence of the dead cone in the forward direction if $M\neq0$.
\item {\it Mid-rapidity $(\eta \sim 0)$}: 
At mid-rapidity ${\cal R}$ depends only weakly on $\eta$ as
\begin{eqnarray}
{\cal R}_{\eta \sim 0}
&\rightarrow & 12g^2\frac{1}{\omega^2} \left (1+\frac{M^2}{s}\right )^{-2} \
\left[1 - 4\eta \frac{M^2}{s+M^2}\right] \,.
\end{eqnarray}

\item {\it Backward rapidity $(\eta \ll 0)$}:
Here (\ref{ratio_R}) becomes
\begin{eqnarray}
{\cal R}_{\eta \ll 0}
& \rightarrow & 3g^2\frac{1}{\omega^2}e^{-2\eta}  
={\cal R}_{\eta \ll 0}^{M\rightarrow 0}  \,.
\end{eqnarray}
In this region the gluon emission does not depend on the mass and is, therefore, the same for heavy as well as light quarks. This is an important aspect for gluon emission off a heavy quark \cite{mgm02,dokshit01,dead1,alam10,kope10,Gossiaux10}.
\end{enumerate}

We also note the dominant process (i.e., $Qg\rightarrow Qgg$) where a gluon acts as a target. Within the hierarchy (\ref{hierarchy}) it differs from  $Qq\rightarrow Qqg$ only by a color Casimir factor $C_A/C_F=9/4$ as
\begin{eqnarray}
\left|{\cal M}_{Qg\rightarrow Qgg}\right|^2=\frac{C_A}{C_F}\left|{\cal M}_{Qq\rightarrow Qqg}\right|^2~,
\end{eqnarray}
since the two body part is given as
\begin{eqnarray}
\left|{\cal M}_{Qg\rightarrow Qg}\right|^2=\frac{C_A}{C_F}\left|{\cal M}_{Qq\rightarrow Qq}\right|^2~,
\end{eqnarray}
 and the other factors are the same for both processes in the considered approximations. Therefore, the factors ${\cal D}$ and $\cal R$ remain unchanged.

\begin{figure}[t]
\includegraphics[width=0.8\linewidth]{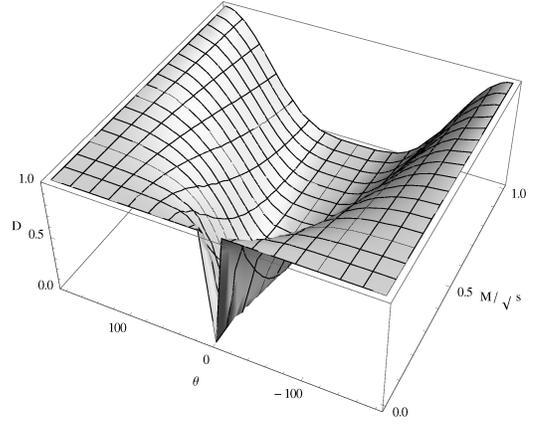}
\caption{The suppression factor ${\cal D}$ from (\ref{new_dead_cone}) as a function of $\theta$ and $M/\sqrt{s}$.}
\label{3D_plot}
\end{figure}
In Fig.~\ref{3D_plot} the suppression factor ${\cal D}$
[cf. Eq.~\eqref{new_dead_cone}] is plotted as a function of $\theta$ and $M/{\sqrt{s}}$.
Around $\theta  \ll 1$ we observe a canyon for small $M/{\sqrt{s}}$ and a valley for large $M/{\sqrt{s}}$, which clearly indicate a presence of a dead cone in the forward direction with respect to the propagating heavy quark. The spread of the dead cone increases as $M/{\sqrt{s}}$ increases. In the backward region, $\theta \sim \pm \pi$, the suppression factor saturates to unity. This suggests that the quark mass plays only a role in the forward direction when the energy of the quark becomes of the order of its mass.

The possibility of this large angle scattering might be important for heavy-ion phenomenology in the context of the non-photonic electron data at RHIC and LHC. Furthermore, it might also have an impact on the description of the forward-backward asymmetry of dijets and the seen energy deposition at large angles in respect to the leading jet \cite{CMS}.

\begin{figure}[t]
\includegraphics[width=0.8\linewidth]{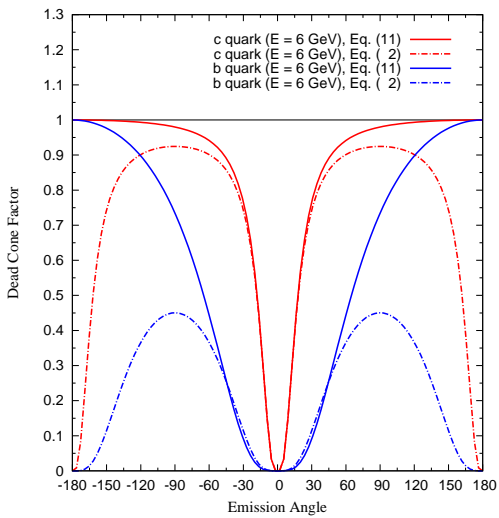}
\caption{(color online) Comparison of our result Eq.~\eqref{new_dead_cone} for the suppression factor ${\cal D}$ to Eq.~\eqref{old_dead_cone_sin2} as a function of $\theta$ for charm and bottom quarks with energy $E= 6 \, {\rm GeV}$. As a note, a mass of $1.2 \, {\rm GeV}$ ($4.2 \, {\rm GeV}$) for  charm (bottom) quarks has been used.}
\label{2D_plot}
\end{figure}

Figure~\ref{2D_plot} compares our result for the generalized suppression factor ${\cal D}$ in (\ref{new_dead_cone}) to that given in (\ref{old_dead_cone_sin2}) as a function of the emission angle ${\theta}$ for charm and bottom quarks with $E= 6 \, {\rm GeV}$. Equation~(\ref{old_dead_cone_sin2}) agrees with our result in the domain of a small emission angle. However, the little variation in this region is due to the constrain $M \ll \sqrt{s}$ employed in earlier calculations, whereas no such constrain is set in our calculation. In contrast to (\ref{old_dead_cone_sin2}) our result for the suppression factor ${\cal D}$ approaches to unity for large emission angles. This indicates that the backward emission is as strong as for light quarks.

In summary, we derived a compact expression that contains a generalized suppression factor for gluon emission off a heavy quark through the scattering with a light parton. In the appropriate limit this expression reduces to the usually known dead cone factor. Our analysis shows that there is a suppression of soft gluon emission due to the mass of the heavy quark in the forward direction. On the other hand, the present findings also indicate that a heavy quark emits a soft gluon almost similar to that of a light quark in the backward rapidity region. This result might have important consequences for a better understanding of heavy flavor energy loss in heavy ion collisions.

\begin{acknowledgments}
This work was supported by the Helmholtz International Center for FAIR within the framework of the LOEWE program launched by the State of Hesse.
M.M. is supported by Ministerio de Ciencia e Innovacion of Spain
under project FPA2009-06867-E. R.A. is grateful to Lab Saha and Satyajit Seth.
\end{acknowledgments}

\end{document}